\documentclass[11pt]{article}
\usepackage[pctex32]{graphics}

\def\ben{\begin{equation}}
\def\een{\end{equation}}

\def\nn{\nonumber} \def\bd{\begin{document}} \def\ed{\end{document}}
\def\ds{\documentstyle} \let\fr=\frac \let\bl=\bigl \let\br=\bigr
\let\Br=\Bigr \let\Bl=\Bigl
\let\bm=\bibitem
\let\na=\nabla
\let\pa=\partial \let\ov=\overline
\newcommand{\be}{\begin{equation}}
\newcommand{\ee}{\end{equation}}
\def\ba{\begin{array}}
\def\ea{\end{array}}
\def\ft#1#2{{\textstyle{\frac{\scriptstyle #1}{\scriptstyle #2} } }}
\def\fft#1#2{{\frac{#1}{#2}}}
\def\del{\partial}
\def\vp{\varphi}
\def\sst#1{{\scriptscriptstyle #1}}
\def\oneone{\rlap 1\mkern4mu{\rm l}}
\def\td{\tilde}
\def\wtd{\widetilde}
\def\ie{{\it i.e.\ }}
\def\dalemb#1#2{{\vbox{\hrule height .#2pt
        \hbox{\vrule width.#2pt height#1pt \kern#1pt
                \vrule width.#2pt}
        \hrule height.#2pt}}}
\def\square{\mathord{\dalemb{6.8}{7}\hbox{\hskip1pt}}}
\newcommand{\ho}[1]{$\, ^{#1}$}
\newcommand{\hoch}[1]{$\, ^{#1}$}
\newcommand{\bea}{\setlength\arraycolsep{2pt} \begin{eqnarray}}
\newcommand{\eea}{\end{eqnarray}}
\newcommand{\ra}{\rightarrow}
\newcommand{\lra}{\longrightarrow}
\newcommand{\Lra}{\Leftrightarrow}
\newcommand{\bp}{\tilde \beta^\prime}
\newcommand{\tr}{{\rm tr} }
\newcommand{\Tr}{{\rm Tr} }
\def\0{{\sst{(0)}}}
\def\1{{\sst{(1)}}}
\def\2{{\sst{(2)}}}
\def\3{{\sst{(3)}}}
\def\4{{\sst{(4)}}}
\def\5{{\sst{(5)}}}
\def\6{{\sst{(6)}}}
\def\7{{\sst{(7)}}}
\def\8{{\sst{(8)}}}
\def\m{{\sst{(m)}}}
\def\n{{\sst{(n)}}}
\def\cA{{{\cal A}}}
\def\cB{{{\cal B}}}
\def\cF{{{\cal F}}}
\def\cG{{{\cal G}}}
\def\cH{{{\cal H}}}
\def\tV{\widetilde V}
\def\tW{\widetilde W}
\def\tH{\widetilde H}
\def\tE{\widetilde E}
\def\tF{\widetilde F}
\def\tA{\widetilde A}
\def\im{{{\rm i}}}
\def\tY{{{\wtd Y}}}
\def\ep{{\epsilon}}
\def\vep{{\varepsilon}}
\def\bD{{{\bar D}}}
\def\R{{{\mathbb R}}}
\def\C{{{\mathbb C}}}
\def\H{{{\mathbb H}}}
\def\CP{{{\mathbb C}{\mathbb P}}}
\def\RP{{{\mathbb R}{\mathbb P}}}
\def\Z{{{\mathbb Z}}}
\def\bA{{{\mathbb A}}}
\def\bB{{{\mathbb B}}}
\def\bC{{{\mathbb C}}}
\def\bD{{{\mathbb D}}}
\def\bE{{{\mathbb E}}}
\def\bZ{{{\mathbb Z}}}
\def\Re{{{\frak{Re}}}}
\def\Im{{{\frak{Im}}}}
\def\cosec{{\,\hbox{cosec}\,}}
\def\Gm{{\Gamma_{\!\! -}}}
\def\Gp{{\Gamma_{\!\! +}}}
\def\stan{{standard }}
\def\nonstan{{supernumerary }}
\def\p{{\partial}}
\def\kdel#1{{\fft{\del}{\del#1}}}

\def\bog{{Bogomolny }}
\def\om{{\omega}}

\newcommand{\nnr}{\nonumber \\}
\newcommand{\pd}{\partial}
\newcommand{\ud}{\textrm{d}}
\newcommand{\dTH}{T^{\prime \, 0}_\textrm{H}}
\newcommand{\dOi}{\Omega^{\prime \, 0}_i}
\newcommand{\bx}{{\bf x}}

\begin{document}

\vspace{5mm}
\begin{center}
{\Large \bf Extremal black holes in the Ho\v{r}ava-Lifshitz
gravity } \vspace{12mm}

{\large   Hyung Won Lee, Yong-Wan Kim, and Yun Soo Myung
\footnote{e-mail
 address: ysmyung@inje.ac.kr}}
 \\
\vspace{10mm} {\em Institute of Basic Science and School of
Computer Aided Science \\ Inje University, Gimhae 621-749, Korea}
\end{center}

\begin{center}

\underline{Abstract}
\end{center}

  We study the near-horizon geometry of extremal black holes
in the $z=3$ Ho\v{r}ava-Lifshitz gravity with a flow parameter
$\lambda$. For $\lambda>1/2$,  near-horizon geometry of extremal
black holes are AdS$_2 \times S^2$ with different radii, depending
on the (modified) Ho\v{r}ava-Lifshitz gravity. For $1/3\le \lambda
\le 1/2$, the radius $v_2$ of $S^2$ is negative, which means that
the near-horizon geometry is ill-defined and  the corresponding
Bekenstein-Hawking entropy is zero. We show explicitly that the
entropy function approach does not work for obtaining the
Bekenstein-Hawking entropy of extremal black holes.

\vspace{15pt}

\thispagestyle{empty}





\newpage
\section{Introduction}
Recently Ho\v{r}ava has proposed a renormalizable theory of
gravity at a Lifshitz point~\cite{ho1},  which  may be regarded as
a UV complete candidate for general relativity. At short distances
the theory of  Ho\v{r}ava-Lifshitz (HL) gravity describes
interacting nonrelativistic gravitons and is supposed to be power
counting renormalizable in (1+3) dimensions. Recently, the HL
gravity theory has been intensively investigated in~\cite{ho2},
 its cosmological applications in
~\cite{cos1,cos2}, and its black hole solutions in
~\cite{LMP,CCO1,CLS,CY,MK,KS,CCO2,Gho,Myung,CJ1,CJ2,CWa,park,BGS,GH,CL,PW,CYZ,CP,Majhi,CW,LHK,WLWG,MyungL,KPPT}.

There are two classes of Ho\v{r}ava-Lifshitz gravity in the
literature:  with/without the projectability condition where the
former (latter) implies that the lapse function  depends on time
(time and space).

Concerning the spherically symmetric solutions without the
projectability condition, L\"u-Mei-Pope (LMP) have obtained the
black hole solution with a flow parameter $\lambda$~\cite{LMP} and
topological black holes were found in \cite{CCO1}. Its
thermodynamics were studied in \cite{MK,CCO2} but there remain
unclear issues in defining the ADM mass and entropy because its
asymptotes are Lifshitz   for $1/3\le \lambda < 3(0<z \le 4)$. On
the other hand, Kehagias and Sfetsos (KS) have found the $\lambda=1$
black hole solution in asymptotically flat spacetimes using the
modified HL gravity without the projectability condition~\cite{KS}.
Its thermodynamics was defined in~\cite{Myung} but recently, the
entropy was argued to take the Bekenstein-Hawking
form~\cite{myungent,WJDC}. Park has obtained a $\lambda=1$ black
hole solution with two parameter $\omega$ and
$\Lambda_W$~\cite{park}. Within the projectable theories, their
black hole solutions are less interesting~\cite{GPW}.

Before proceeding, we wish to point out that the Ho\v{r}ava black
holes (LMP and KS black holes) are completely different from the IR
black holes (Schwarzschild-AdS and Schwarzschild black holes). The
Ho\v{r}ava  black holes have their extremal black holes, whereas the
IR black holes do not have extremal black holes. In the Ho\v{r}ava
black holes, the charge-like quantities (pseudo charge) are related
to the cosmological constant $-\frac{1}{\Lambda_W}$ in the LMP black
holes and parameter $\frac{1}{2\omega}$ in the KS black hole. This
feature is very special, in compared to the Reissner-Norstr\"om-AdS
and Reissner-Norstr\"om black holes with the electric (magnetic)
charge ``$Q$" which were obtained from the relativistic theories. We
wish to mention that in the non-relativistic theories, the notion of
horizon, temperature, and entropy was not well-defined~\cite{KKbh}.
 Although most solutions have  horizons,
 a part of solutions appears to be horizonless for particles with ultra-luminal dispersion relations.

In order that the thermodynamics of an extremal black hole is
explored completely, the whole spacetimes should be  known,
including the near-horizon geometry and  asymptotic structures.
However, for $1/3\le \lambda <3$, the LMP black holes have
asymptotically Lifshitz which is not yet understood fully. In this
sense, we call these Lifshitz black holes. Thus, it is still lack
for understanding thermodynamics of the Lifshitz black holes because
the conserved quantities are not defined unambiguously.

 In this work, we investigate the near-horizon geometry (AdS$_2\times
 S^2$) of the extremal black holes
to explore the unknown solution in asymptotically flat spacetimes
and to understand  the Lifshitz black holes.   For this purpose, we
obtain AdS$_2\times S^2$ by exploring  the KS solution and the LMP
solution. Then, we compare these with  curvature radius $v_1$ of
AdS$_2$ and curvature radius $v_2$  of $S^2$ obtained by solving
full Einstein equations on the AdS$_2 \times S^2$ background
directly. For $\lambda>1/2$ case, the  near-horizon geometry of
extremal black holes are AdS$_2 \times S^2$ with different
curvatures, depending on the (modified) Ho\v{r}ava-Lifshitz gravity.
For $1/3\le \lambda < 1/2$ case, the radius $v_2$ of $S^2$ is always
negative and thus the corresponding near-horizon geometry is
ill-defined. Hence, this case will be ruled out as candidate for the
extremal black holes in the  Ho\v{r}ava-Lifshitz gravity.

 \section{HL gravity}
Introducing the ADM formalism where the metric is parameterized
\cite{adm}
\be ds_{ADM}^2= - N^2  dt^2 + g_{ij} \Big(dx^i - N^i dt\Big)
\Big(dx^j - N^j dt\Big)\,, \ee
the Einstein-Hilbert action can be expressed as
\be \label{Eins} S_{EH} = \fft{1}{16\pi G} \int d^4x \sqrt{g} N
\Big[K_{ij} K^{ij} - K^2 + R - 2\Lambda\Big], \ee
where $G$ is Newton's constant and extrinsic curvature $K_{ij}$
takes the form
\be K_{ij} = \fft{1}{2N} \Big(\dot g_{ij} - \nabla_i N_j -
\nabla_j N_i\Big)\,. \ee
Here, a dot denotes a derivative with respect to $t$. An action of
the non-relativistic renormalizable gravitational theory  is given
by~\cite{KS} \be S_{HL}=\int dtd^3\bx \Big[{\cal L}_K + {\cal
L}_V\Big],  \label{action1} \ee where the kinetic term is  given by
\be {\cal L}_K =\frac{2}{\kappa^2}\sqrt{g} N K_{ij}{\cal
G}^{ijkl}K_{kl}= \frac{2}{\kappa^2}\sqrt{g}
N\Big(K_{ij}K^{ij}-\lambda K^2\Big), \ee with the DeWitt metric
 \be {\cal G}^{ijkl}=
\frac{1}{2}\Big(g^{ik}g^{jl}-g^{il}g^{jk}\Big)-\lambda
g^{ij}g^{kl} \ee
 and its inverse metric
 \be {\cal
G}_{ijkl}=\frac{1}{2}\Big(g_{ik}g_{jl}-g_{il}g_{jk}\Big)-\frac{\lambda}{3\lambda-1}g_{ij}g_{kl}.\ee

The potential term is determined by the detailed balance condition
(DBC) as \bea {\cal L}_V=-\frac{\kappa^2}{2}\sqrt{g}N E^{ij}{\cal
G}_{ijkl}E^{kl}&=&
\sqrt{g}N\Bigg\{\frac{\kappa^2\mu^2}{8(1-3\lambda)}\Big(\frac{1-4\lambda}{4}R^2
+\Lambda_WR-3\Lambda_W^2\Big)\nn \\
 &-&\frac{\kappa^2}{2\eta^4} \left(C_{ij}
-\frac{\mu \eta^2}{2}R_{ij}\right) \left(C^{ij} -\frac{\mu
\eta^2}{2}R^{ij}\right) \Bigg\}.\label{action2} \eea Here the $E$
tensor is defined by \be E^{ij}=\frac{1}{\eta^2}
C^{ij}-\frac{\mu}{2} \Big(R^{ij}-\frac{R}{2}
g^{ij}+\Lambda_Wg^{ij}\Big) \ee with the Cotton tensor $C_{ij}$ \be
C^{ij}=\frac{\epsilon^{ik\ell}}{\sqrt{g}}\nabla_k\left(R^j{}_\ell
-\frac14R\delta_\ell^j\right).\label{def.K.C} \ee Explicitly,
$E_{ij}$ could be derived  from the Euclidean topologically massive
gravity \be E^{ij}=\frac{1}{\sqrt{g}} \frac{\delta W_{TMG}}{\delta
g_{ij}} \ee with \be W_{TMG}=\frac{1}{\eta^2} \int d^3 x
\epsilon^{ijk}\Big(\Gamma^m_{il}\partial_j \Gamma^l_{km}+\frac{2}{3}
\Gamma^n_{il} \Gamma^l_{jm} \Gamma^m_{kn} \Big)- \mu \int d^3x
\sqrt{g}(R-2\Lambda_W), \ee where $\epsilon^{ikl}$ is a tensor
density with $\epsilon^{123}=1$. In the IR limit,  comparing
Eq.(\ref{action1}) with Eq.(\ref{Eins}) of general relativity, the
speed of light, Newton's constant and the cosmological constant are
given by
\be c=\fft{\kappa^2\mu}{4}
\sqrt{\fft{\Lambda_W}{1-3\lambda}}\,,\qquad
G=\fft{\kappa^2}{32\pi\,c}\,,\qquad \Lambda=\ft32
\Lambda_W\,.\label{cg} \ee
The equations of motion were derived in \cite{cos1} and \cite{LMP}.
We would like to mention that the IR vacuum of this theory is anti
de Sitter (AdS$_4$) spacetimes. Hence, it is interesting to take a
limit of the theory, which may lead to  a Minkowski vacuum in the IR
sector. To this end, one may deform the theory by introducing a
modified term of ``$\mu^4R$" $(\tilde{{\cal L}}_V={\cal
L}_V+\sqrt{g}N \mu^4R)$ and then, take the $\Lambda_W \to 0$ limit.
This does not alter the UV properties of the theory, while it
changes the IR properties. That is, there exists a Minkowski vacuum,
instead of an AdS vacuum. In the IR limit,   the speed of light and
Newton's constant are given by
\be c=\fft{\kappa^2\mu^4}{2},~ G=\fft{\kappa^2}{32\pi\,c},
~\lambda=1.\label{kh} \ee

Taking $N^i=0$ and $K_{ij}=0$, a spherically symmetric solution
could be obtained with the metric ansatz
\begin{eqnarray}
\label{ssm} ds_{SS}^2 = - N^2(\rho)\,d\tau^2 +
\frac{d\rho^2}{f(\rho)} + \rho^2 (d\theta^2 +\sin^2\theta
d\phi^2).
\end{eqnarray}
Substituting the metric ansatz (\ref{ssm}) into $\tilde{{\cal L}}_V$
with $R=-\frac{2}{\rho^2}(\rho f'+f-1)$, one has the reduced
Lagrangian
\begin{eqnarray}
\label{react} \tilde{{\cal L}}^{SS}_V&=&\frac{ \kappa^2\mu^2 N
}{8(1-3\lambda)\sqrt{f}}\Bigg[  \frac{\lambda-1}{2} f'^2 -
\frac{2\lambda (f-1)}{ \rho}f' \\ \nn &+&
\frac{(2\lambda-1)(f-1)^2}{ \rho^2}-2(w-\Lambda_W)(1 - f - \rho
f')-3 \Lambda_W^2\rho^2\Bigg].
\end{eqnarray} Here the parameter $\omega$   controls the UV effects
\be w=\frac{8\mu^2(3\lambda-1)}{\kappa^2},\ee which is positive for
$\lambda>1/3$.
 This is possible because the
metric ansatz shows all the allowed singlets which are compatible
with the $SO(3)$ action on the $S^2$. In other words, the ADM
decomposition of HL gravity implies naturally the presence of a
spherically symmetric static solution, in addition to  the
time-evolving solution of hypersurfaces for foliation preserving
diffeomorphisms~\cite{ho1}.

\section{Kehagias-Sfetsos black hole}
 For
$\lambda=1~(w=16\mu^2/\kappa^2\equiv 1/2Q^2)$ and $\Lambda_W=0$,
we have Kehagias-Sfetsos (KS) black hole solution where $f$ and
$N$ are determined to be
\begin{eqnarray}
\label{sol1}
N^2=f=\frac{2(\rho^2-2M\rho+Q^2)}{\rho^2+2Q^2+\sqrt{\rho^4+8Q^2
M\rho}}.
\end{eqnarray}
Here $M$ may be related   to the ADM mass. It seems that the metric
function $f$ looks like that of Reissner-Nordstr\"om (RN) black
hole.  The outer (event) and inner (Cauchy) horizons are given by
\begin{eqnarray}
\rho_\pm=M\pm \sqrt{M^2-Q^2}
\end{eqnarray}
which is  the same form as in the RN black hole.  The extremal black
hole is located at
\begin{equation}
\label{extremal} \rho_\pm=\rho_e = M=Q.
\end{equation}
This is why we take $\frac{1}{2\omega}$ to be the charge-like
quantity (pseudo charge) $Q^2$. In the limit of $ Q^2 \to 0 (\omega
\to \infty)$, we recover the Schwarzschild black hole as the IR
black hole.

However,  for large $\rho$, we approximate the metric function as
\be f \to 1-\frac{2M}{\rho}+\frac{4M^2Q^2}{\rho^4}+\cdots, \ee which
is different from the RN-metric function \be f_{RN} =
1-\frac{2M}{\rho}+\frac{Q^2}{\rho^2}.\ee

At this stage,  we  define the Bekenstein-Hawking entropy for the
extremal KS black hole\footnote{At present, we may have two kinds of
entropy: one is the logarithmic entropy of
$S=\frac{A}{4}+\frac{\pi}{\omega}\ln\Big[\frac{A}{4}\Big]$~\cite{CL,myungbhent}
and the other is the Bekenstein-Hawking entropy
$S_{BH}=\frac{A}{4}$~\cite{myungent,WJDC}. The former can be
obtained from the first law of thermodynamics $dM=T_HdS$ provided
$M$ and $T_H$ are known. The temperature is defined from the surface
gravity at the horizon and thus, it is independent of the asymptotic
structure. On the other hand, the ADM  mass $M$ depends on the
asymptotic structure because it belongs to a conserved quantity
defined at infinity. Hence, the entropy also depends on the mass
$M$. However, it is hard to accept the logarithmic entropy without
considering quantum or thermal corrections. Therefore, it would be
better to use the area-law of the Bekenstein-Hawking entropy to
derive the ADM mass using the first law. In this work, we choose the
Bekenstein-Hawking entropy as the entropy of the Ho\v{r}ava black
holes.  } as
\begin{equation}
\label{bh-entropy} S_{\rm BH} = \frac{A}{4} = \pi Q^2.
\end{equation}
In order to explore the near-horizon geometry $AdS_2 \times S^2$ of
extremal black hole, we introduce new coordinates $t$ and $r$ using
relations
\begin{equation}
\label{coord-trans} t = \frac{\epsilon}{3Q^2} \tau, \,\,\,\, r =
\frac{\rho - Q}{\epsilon},
\end{equation}
where $\epsilon$ is an arbitrary constant.  Then, the extremal
solution can be expressed by  new coordinates as
\begin{eqnarray}
\label{ext-metric} ds^2_{\rm ext} &=& -\frac{2 r^2}{(Q+\epsilon
r)^2 + 2Q^2 + \sqrt{(Q+\epsilon r)^4 +
 8Q^3 (Q+\epsilon r)}} 9Q^4 dt^2 \nonumber \\
&& + \frac{(Q+\epsilon r)^2 + 2Q^2 + \sqrt{(Q+\epsilon r)^4 + 8Q^3
(Q+\epsilon r)}}{2 r^2} dr^2 + (Q+\epsilon r)^2 d\Omega^2_2.
\end{eqnarray}
Taking  the `near-horizon' limit of $\epsilon \rightarrow 0$ leads
to \be \label{ads-metric} ds^2_{\rm KS} =v_1 \left ( -r^2 dt^2 +
\frac{dr^2}{r^2} \right ) + v_2 d\Omega^2_2 \ee with \be
v_1=3Q^2,~~v_2=Q^2. \ee This shows   clearly  $AdS_2 \times S^2$ in
the Poincare coordinates. Its curvature is given by \be ^{(4)}R_{\rm
KS}=R_{AdS_2}+R_{S^2}=-\frac{2}{3Q^2}+\frac{2}{Q^2}.\ee In this
case, the KS black hole solution interpolates between  $AdS_2\times
S^2$ in the near-horizon geometry and Minkowski spacetimes at
asymptotic infinity.

However, the RN  black hole has the Bertotti-Robinson metric as its
near-horizon geometry
 \be \label{RN-metric} ds^2_{\rm RN} =Q^2 \left ( -r^2 dt^2 +
\frac{dr^2}{r^2} \right ) + Q^2 d\Omega^2_2.\ee Its curvature is
given by \be ^{(4)}R_{\rm RN}=-\frac{2}{Q^2}+\frac{2}{Q^2}.\ee

Finally, we mention that the general solution with $\lambda$ is not
found in asymptotically flat spacetimes because  third-order
derivatives make it difficult to solve the Einstein equation.

\section{L\"u-Mei-Pope black hole solution}
In this case, we take  $\omega=0$ in Eq. (\ref{react}) by dropping
the modified term of $\mu^4R$. The L\"{u}-Mei-Pope (LMP) solutions
for $z=3$ Ho\v{r}ava-Lifshitz gravity are given by \be f(x) = 1 +
x^2 - \alpha x^{p_\pm(\lambda)},~~ N(x) = x^{q_\pm(\lambda)}
\sqrt{f(x)}, \label{bh_sol} \ee where \be x = \sqrt{-\Lambda_W} r,~
p_\pm(\lambda) = \frac{2 \lambda \pm \sqrt{6 \lambda -2}}{\lambda
-1},~~q_\pm(\lambda) =- \frac{1 + 3 \lambda \pm 2 \sqrt{6 \lambda
-2}}{\lambda -1}.\ee
 In this work, we choose
$p_-(\lambda)=p(\lambda)$ and $q_-(\lambda)=q(\lambda)$ only. Its
extremal black hole with $f(x_e) =0$ and $ f'(x_e) = 0$ are located
as \be \label{extrecon} x_e=0,~{\rm for}~\frac{1}{3}\le \lambda \le
\frac{1}{2};~~
x_e=\sqrt{\frac{p(\lambda)}{2-p(\lambda)}}=\sqrt{\frac{2\lambda-\sqrt{6\lambda-2}}{-2+\sqrt{6\lambda-2}}},~{\rm
for}~\lambda
> \frac{1}{2}. \ee
The Bekenstein-Hawking entropy for the extremal LMP black holes is
defined by \be S_{\rm BH}=\pi x^2_e\Big[\frac{1}{-\Lambda_W}\Big],
\ee where $-\frac{1}{\Lambda_W}$ is interpreted as a pseudo charge.
Hence the non-zero  entropy is available for $\lambda>1/2$ only. The
metric function $f(x)$ can be expanded around extremal point to find
its near-horizon geometry AdS$_2 \times S^2$ as
\begin{equation}
f(x) \approx \frac{f''(x_e)}{2} (\epsilon \rho)^2 + {\cal
O}(\epsilon^3), \label{expand}
\end{equation}
where $x - x_e = \epsilon \rho$ and $\tilde t = \epsilon A t$ with
\be A = \frac{f''(x_e)x_e^{q(\lambda)}}{2} \sqrt{-\Lambda_W}. \ee
The line element is given by~\cite{GH}
\begin{equation}
ds^2 =  \frac{2}{f''(x_e)} \Big[\frac{1}{-\Lambda_W}\Big] \left ( -
\rho^2 d{\tilde t}^2 + \frac{d\rho^2}{\rho^2} \right ) +
\Big[\frac{x^2_e}{-\Lambda_W}\Big] d\Omega^2 \label{metric_ads_LH1}
\end{equation}
with \be f''(x_e) =  4 - 2 p(\lambda) . \label{ddf_x0} \ee
Comparing the above expression with AdS$_2 \times S^2$, we obtain
\begin{eqnarray}
v_1 &=& \frac{1}{2 - p(\lambda)} \Big[\frac{1}{-\Lambda_W}\Big] = \frac{v_2}{p(\lambda)}, \label{v1} \\
v_2 &=& \frac{p(\lambda)}{2 -
p(\lambda)}\Big[\frac{1}{-\Lambda_W}\Big]. \label{v2}
\end{eqnarray}
We express the Bekenstein-Hawking entropy as $S_{\rm BH}=\pi v_2$.
For $\lambda=1$, we obtain  the LMP case \be x_e^2  =\frac{1}{3}, ~~
v_1 = \frac{2}{3} \Big[\frac{1}{-\Lambda_W}\Big] = 2 v_2,~~ v_2 =
\frac{1}{3} \Big[\frac{1}{-\Lambda_W}\Big]. \ee Importantly, from
Fig. 1, it is problematic to define the near-horizon geometry of
extremal black hole for $1/3\le \lambda \le 1/2$ because $v_2$ is
negative for $1/3\le \lambda < 1/2$~($p(\lambda)<0$) and $v_2$ is
zero for $\lambda=1/2~(p(\lambda)=0)$. That is, its near-horizon
geometry is ill-defined. In addition, the  Bekenstein-Hawking
entropy is zero.

It seems that  the Bertotti-Robinson geometry (RN black hole) of
$v_1=v_2=-1/\Lambda_W$ is recovered from $\lambda=3$ LMP black hole
with $x_e^2=1$ \be f(x)=1+x^2-\alpha x,~~N(x)=\frac{\sqrt{f(x)}}{x}
.\ee However, its asymptotic spacetimes are completely different
from asymptotically flat spacetimes  of the RN black hole.

\begin{figure}[t!]
   \centering
   \includegraphics{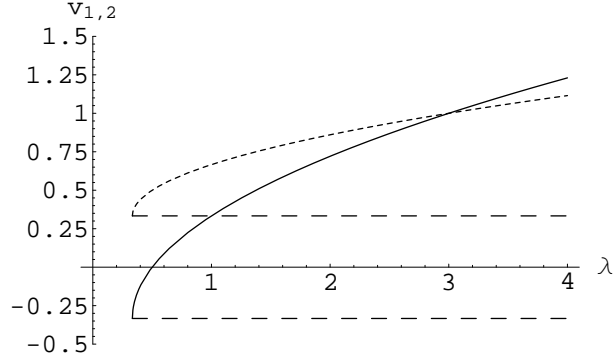}
\caption{$v_1(\lambda)$ (dotted curves) and $v_2(\lambda)$ (solid
curves) graphs with $\Lambda_W=-1$. We find
$v_1(1/3)=1/3,~v_1(1/2)=1/2,~v_1(1)=2/3$, and $v_1(3)=1$, while we
observe $v_2(1/3)=-1/3,~v_2(1/2)=0,~v_2(1)=1/3$, and $v_2(3)=1$. }
\label{fig.1}
\end{figure}
For $1/2<\lambda<3$, the LMP black hole solutions interpolate
between   AdS$_2 \times S^2$ with $v_1>v_2$  in  the near-horizon
geometry of extremal black hole and Lifshitz at asymptotic infinity,
while for $\lambda>3$, these interpolate  AdS$_2 \times S^2$ with
$v_1<v_2$ in the near-horizon and Minkowski spacetimes at asymptotic
infinity.

\section{Solution to full equations on AdS$_2 \times S^2$}
In this section, we wish to  find  $v_1$-and $ v_2$-solutions by
solving full Einstein  equations on the AdS$_2 \times S^2$
background with ${\cal L}_K=0$. This study will be very useful for a
further work on Ho\v{r}ava black holes because it may provide a hint
to explore the unknown black hole solutions.

A variation to $\tilde{{\cal L}}_V$ with respect to $N$ is modified
as \bea \frac{\delta \tilde{{\cal L}}_V}{\delta N}=0 :
\sqrt{g}\Bigg[
\frac{\kappa^2\mu^2(\Lambda_W-\omega)}{8(1-3\lambda)}R&-&\frac{3\kappa^2\mu^2\Lambda_W^2}{8(1-3\lambda)}
\\ \nn
&+&
\frac{\kappa^2\mu^2(1-4\lambda)}{32(1-3\lambda)}R^2-\frac{\kappa^2}{2\eta^4}
\Big(C_{ij}-\frac{\mu \eta^2}{2}R_{ij}\Big)^2\Bigg]=0. \eea A
variation to
 with respect to the shift function  $N^j$ is trivial for finding  a black
 hole solution as
 \be
 \frac{\delta \tilde{{\cal L}}_V}{\delta N^i}=0 \to {\rm trivial}.
 \ee
 A variation
to ${\cal L}_K+\tilde{{\cal L}}_V$ with respect to $g_{ij}$ is
changed to \be \frac{\delta \tilde{{\cal L}}_V}{\delta g^{ij}}=0:
{\cal E}_{ij}\equiv
\frac{\kappa^2\mu^2(\Lambda_W-\omega)}{8(1-3\lambda)}E^{(3)}_{ij} +
\frac{\kappa^2\mu^2(1-4\lambda)}{32(1-3\lambda)}E^{(4)}_{ij}
-\frac{\kappa^2\mu}{4\eta^2}E^{(5)}_{ij}-\frac{\kappa^2}{2\eta^4}E^{(6)}_{ij}=0,
\ee where $E_{ij}^{(3)}$ takes the modified form \be
E^{(3)}_{ij}=N\Big(R_{ij}-\frac{1}{2}Rg_{ij}+\frac{3}{2}\frac{\Lambda^2_W}{\Lambda_W-\omega}g_{ij}\Big)
-\Big(\nabla_i\nabla_j-g_{ij}\nabla^2\Big)N \ee and all remaining
terms  are the same as in \cite{LMP}.

Considering AdS$_2 \times S^2$ in Eq. (\ref{ads-metric}), we have
\begin{eqnarray}
N&=&\sqrt{v_1}r,~~g_{ij} = {\rm diag}\Big[\frac{v_1}{r^2}, v_2, v_2
\sin^2
\theta\Big], \nn \\
\label{adsnh}R_{ij} &=& {\rm diag}\Big[0, 1, \sin^2 \theta\Big],
~C_{ij} = {\rm diag}\Big[0, 0,0\Big],~K_{ij} = {\rm diag}\Big[0, 0,0\Big] \\
R &=& \frac{2}{v_2},~~K=0. \nn
\end{eqnarray}
The $N$-variation leads to the equation obtained when varying with
respect to $v_1$ as
\begin{equation}
\frac{\kappa^2\mu^2\sqrt{v_1}\sin \theta}{8(3 \lambda -1)
v_2r}\Bigg[3 \Lambda_W^2 v_2^2 +2 (\omega-\Lambda_W) v_2 - 2 \lambda
+ 1\Bigg] = 0. \label{eq_N_MHL}
\end{equation}

The  $g_{ij}$-variation  leads to three component equations
\begin{eqnarray}
{\cal E}_{rr} &=& \frac{\kappa^2 \mu^2 v_1^{3/2}}{16(3 \lambda -1)
v_2^2 r}
\Bigg[3 \Lambda_W^2 v_2^2 +2(\omega - \Lambda_W) v_2 - 2 \lambda + 1\Bigg] = 0, \label{eq_err_MHL}\\
{\cal E}_{\theta\theta} &=& \frac{\kappa^2 \mu^2 r}{16(3 \lambda
-1) v_1^{1/2} v_2} \Bigg[ v_1 \Big(3 \Lambda_W^2 v_2^2 + 2 \lambda
- 1 \Big)-2 \Big( (\omega-\Lambda_W) v_2 + \lambda \Big)
 v_2 \Bigg] = 0, \label{eq_ethth_MHL}\\
{\cal E}_{\phi\phi} &=& \sin^2 \theta  {\cal E}_{\theta\theta} =
0.\label{eq_ephiphi_MHL}
\end{eqnarray}
Eqs. (\ref{eq_N_MHL}) and (\ref{eq_err_MHL}) give the same
equation for $v_2$ as
\begin{equation}
3 \Lambda_W^2 v_2^2 + 2(\omega-\Lambda_W) v_2 - 2 \lambda + 1 = 0.
\label{eq_v2_MHL}
\end{equation}
For $\Lambda_W \ne 0$, this equation is solved to give
\begin{equation}
\Lambda_W v_2 = \frac{1}{3}\left ( 1 - \frac{\omega}{\Lambda_W}
\right ) \left ( 1 \pm \sqrt{1 + \frac{6 \lambda -3}{\left ( 1-
\frac{\omega}{\Lambda_W}\right )^2}} \right ).
\label{sol_v2_0_MHL}
\end{equation}
For $\Lambda_W < 0$, the solution is
\begin{equation}
v_2(\lambda,\omega,\Lambda_W) = \frac{1}{3}\left ( 1 -
\frac{\omega}{\Lambda_W} \right ) \left ( \sqrt{1 + \frac{6 \lambda
-3}{\left ( 1- \frac{\omega}{\Lambda_W}\right )^2}} - 1 \right )
\Big[\frac{1}{-\Lambda_W}\Big]. \label{sol_v2_MHL}
\end{equation}
From  Eq. (\ref{eq_ethth_MHL}), we find  for $\Lambda_W \ne 0$
\begin{equation}
v_1(\lambda,\omega,\Lambda_W) = \frac{6 \lambda - 2\left ( 1 -
\frac{\omega}{\Lambda_W} \right )^2
      \left ( 1 - \sqrt{1 + \frac{6 \lambda -3}{\left ( 1- \frac{\omega}{\Lambda_W}\right )^2 }}\right )}
     {6 \lambda - 3 + \left ( 1 - \frac{\omega}{\Lambda_W} \right )^2
     \left ( 1 - \sqrt{1 + \frac{6 \lambda -3}{\left ( 1- \frac{\omega}{\Lambda_W}\right )^2}} \right )^2 }~ v_2(\lambda,\omega,\Lambda_W).
\label{sol_v1_0_MHL}
\end{equation}
Eqs. (\ref{sol_v2_MHL}) and (\ref{sol_v1_0_MHL}) are our main
results.

\subsection{ General KS AdS$_2 \times S^2$-solution}
For $\Lambda_W = 0$, Eq. (\ref{eq_v2_MHL})  becomes
\begin{equation}
2 \omega v_2 -2 \lambda + 1 = 0, \label{eq_N_MHL1}
\end{equation}
which gives the solution
\begin{equation}
v_2 = \frac{2\lambda - 1}{2 \omega}. \label{sol_v2_0}
\end{equation}
From Eq. (\ref{eq_ethth_MHL}) with Eq. (\ref{sol_v2_0}), one has the
relation between $v_1$ and $v_2$ \be \label{sol v2 1}
v_1=\frac{4\lambda-1}{2\lambda-1}v_2. \ee The Bekenstein-Hawking
entropy for  extremal  black holes with $\lambda$ may take \be
S_{\rm BH}=\pi v_2=\pi\Big[\frac{2\lambda-1}{2\omega}\Big]. \ee For
$\lambda=1$, we recover the KS AdS$_2 \times S^2$ exactly as
 \be v_1=3v_2,
~~v_2=\frac{1}{2\omega}=Q^2 \ee whose entropy leads to Eq.
(\ref{bh-entropy}). Although the black hole solution with
arbitrary $\lambda$  is not yet known in asymptotically flat
spacetimes, its near-horizon geometry of extremal black hole with
$\lambda$ could be found from Eqs. (\ref{sol_v2_0}) and (\ref{sol
v2 1}). For $\lambda >1/2$, we can define  extremal black holes
because
 both $v_1$ and $v_2$ are positive.

\subsection{LMP AdS$_2 \times S^2$ solution}
Here we comment on the $\omega=0$ case.  Eq. (\ref{sol_v2_MHL})
provides the solution $v_2$
\begin{equation}
v_2 = \frac{\sqrt{6 \lambda -2}-1}{3}\Big[\frac{1}{-\Lambda_W
}\Big]=\Big[\frac{p(\lambda)}{2-p(\lambda)}\Big]\Big[\frac{1}{-\Lambda_W}\Big].
\label{sol_v2_1}
\end{equation}
From Eq. (\ref{sol_v1_0_MHL}) with $\omega=0$, we obtain
\begin{equation}
v_1 = \frac{\lambda -1 }{2 \lambda  - \sqrt{6 \lambda -2}}
v_2=\frac{v_2}{p(\lambda)}. \label{sol_v1}
\end{equation}
This is the same result as was found from the previous section.
\subsection{General  AdS$_2 \times S^2$ solution}
\begin{figure}[t!]
   \centering
   \includegraphics{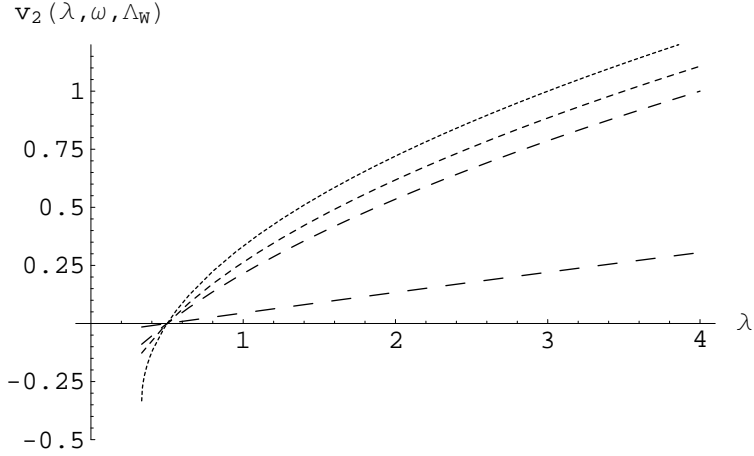}
\caption{$v_2(\lambda,\omega,\Lambda_W)$ graphs with $\Lambda_W=-1$.
We find that $v_2>0,=0,<0$ for
$\lambda>\frac{1}{2},=\frac{1}{2},<\frac{1}{2}$, respectively. Four
graphs show  for $\omega=0,~0.5,~1,$ and 10  from top to bottom.}
\label{fig.2}
\end{figure}
\begin{figure}[t!]
   \centering
   \includegraphics{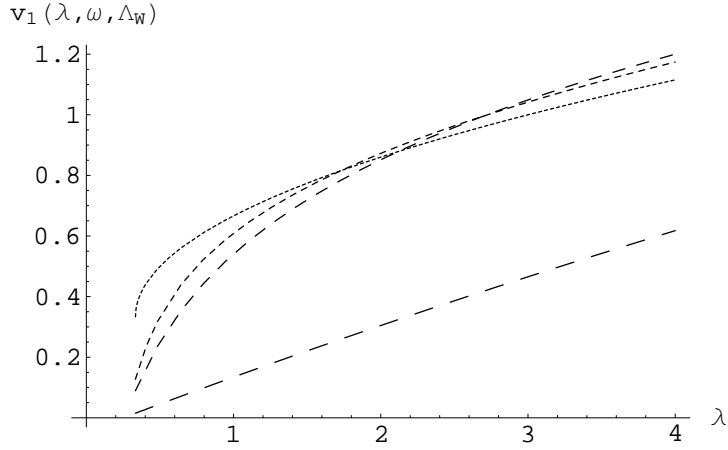}
\caption{$v_1(\lambda,\omega,\Lambda_W)$ graphs with $\Lambda_W=-1$.
It shows that $v_1>0$ for $\lambda>\frac{1}{3}$. Four graphs are for
$\omega=0,~0.5,~1,$ and 10 from top to bottom (along $v_1$-axis). }
\label{fig.3}
\end{figure}
We consider the general case of $\omega\not=0$ and
$\Lambda_W\not=0$.  As is shown in Fig. 2, it is observed  that
$v_2>0$ only for $\lambda>1/2$, irrespective of any values of
$\omega$ and $\Lambda_W<0$. It could be confirmed from the fact
that $v_2=0$ implies $\lambda=1/2$.  We note that the $\lambda=1$
solution describes the near-horizon geometry of Park's
solution~\cite{park}. Also, from Fig. 3, we confirm that $v_1$ is
always positive.

 Finally,  the
entropy of extremal black hole is determined to be  \be S_{\rm
BH}=\pi v_2 \ee where $v_2$ is given by Eq. (\ref{sol_v2_MHL}).

\section{Entropy function approach}
Plugging Eq. (\ref{adsnh}) into  $\tilde{{\cal L}}_V$ leads to the
Lagrangian on AdS$_2\times S^2$
\begin{equation}
\tilde{{\cal L}}^{AdS_2\times S^2}_V =  \frac{\kappa^2 \mu^2v_1 \sin
\theta}{8(3 \lambda -1) v_2}
 \left (
 3 \Lambda_W^2 v_2^2 + 2(\omega - \Lambda_W) v_2 - 2 \lambda +1
 \right ) .
\label{lagrangian-ads}
\end{equation}
which is nothing but \be N \frac{\delta \tilde{\cal L}_V}{\delta
N}\ee in Eq.(\ref{eq_N_MHL}).  After integration over $S^2$, it
takes the form of entropy function~\cite{sen}
\begin{equation}
\tilde{{\cal L}}^{AdS_2}_V =\frac{\pi \kappa^2 \mu^2v_1}{2(3 \lambda
-1) v_2}
 \Bigg[3 \Lambda_W^2 v_2^2 + 2(\omega - \Lambda_W)v_2  -2 \lambda +1
 \Bigg].
\label{lagrangian_ads}
\end{equation}
A variation of $\tilde{{\cal L}}^{AdS_2}_V$ with respect to $v_1$
leads to the known equation (\ref{eq_v2_MHL}) \be \label{ef1}
\frac{\pi \kappa^2 \mu^2}{2(3 \lambda -1) v_2}\Big[3 \Lambda_W^2
v_2^2 + 2(\omega - \Lambda_W) v_2 - 2 \lambda +1\Big]=0, \ee while a
variation with respect to $v_2$ takes a different form \be
3\Lambda_W^2v_2^2+(2\lambda-1)=0. \ee The entropy function
$\tilde{{\cal L}}^{AdS_2}_V$  is zero if  Eq. (\ref{ef1}) is used.
Since this function  contains $v_1$ as a global factor, we cannot
determine  $v_1$ by varying the entropy function with respect to
$v_2$. It seems that this is a handicap in the entropy function
approach to the Ho\v{r}ava  black holes because the HL gravity is a
non-relativistic theory. However, we have determined $v_1$ in Eq.
(\ref{sol_v2_MHL}) when using full Einstein equations on the  AdS$_2
\times S^2$ background.
 Now we wish to clarify  why
the reduced Lagrangian (\ref{lagrangian_ads}) is not  useful  to
determine $v_1$. The entropy function approach is equivalent to
taking into account (\ref{eq_N_MHL}) and (\ref{eq_err_MHL}). On the
other hand, the full Einstein equation on the AdS$_2 \times S^2$
provides two more equations (\ref{eq_ethth_MHL}) and
(\ref{eq_ephiphi_MHL}) than the entropy function approach. These
determine a relation between $v_2$ and $v_1$, as is shown in
Eq.(\ref{sol_v1_0_MHL}).

\section{Discussions}
In this work, we have studied near-horizon geometry AdS$_2 \times
S^2$ of HL black holes. An important relation is $v_1=\frac{v_2}{p}$
and $ v_2=\frac{p}{2-p}[\frac{1}{-\Lambda_W}]$ for the LMP black
holes~\cite{LMP}. Also we have obtained
$v_1=\frac{4\lambda-1}{2\lambda-1}v_2$ and
$v_2=\frac{2\lambda-1}{2\omega}$ for generalized KS black holes. The
$\lambda=1$ case corresponds to the KS solution:
$v_1=3v_2,~v_2=\frac{1}{2\omega}$~\cite{KS}. We regard
$(-\frac{1}{\Lambda_W},\frac{1}{2\omega})$ as pseudo charges as
magnetic charge ``$Q$ " in RN and RN-AdS black holes.

Since $v_2$ is zero for $\lambda=1/2$, we have to classify whole
extremal black holes according to the $\lambda$-value. For
$\lambda>1/2$ case, near-horizon geometry of extremal black holes
are AdS$_2 \times S^2$ with different curvatures, depending on the
(modified) Ho\v{r}ava-Lifshitz gravity. For $1/3 \le \lambda < 1/2$
case, a radius $v_2$ of $S^2$ is negative and thus the corresponding
near-horizon geometry is ill-defined. Hence this case should be
discriminated  from the extremal  black holes in  the
Ho\v{r}ava-Lifshitz gravity. This is clear because as is shown in
Eq.(\ref{extrecon}), the Lifshitz black holes with $1/3 \le \lambda
\le 1/2 (4\le z \le 2)$ have degenerate horizons located at the
origin like massless BTZ black hole in three dimensions.

We hope that the black hole with arbitrary $\lambda$ could be found
soon in asymptotically flat spacetimes  because  its near-horizon
geometry of extremal black hole are known as Eqs. (\ref{sol_v2_0})
and (\ref{sol v2 1}). Also, we expect to obtain a general black hole
whose near-horizon geometry is given by Eqs. (\ref{sol_v2_MHL}) and
(\ref{sol_v1_0_MHL}) soon.

Finally, the entropy function approach does not work  for obtaining
the Bekenstein-Hawking  entropy of extremal Ho\v{r}ava black holes
because the Lorentz-symmetry was broken.

{\it Noted added}--After the appearance of this work in arxiv,
considerable research has indicated that for $1/3 \le \lambda \le
1$,  the projectable version of HL gravity has a serious problem
called as strong coupled problem. In the projectable theories, the
authors~\cite{SVW,BPS} have first argued that $\psi$ is propagating
around the Minkowski space but it has a negative kinetic term,
showing a ghost instability. In this case, the Ho\v{r}ava scalar
becomes ghost if the sound speed square ($c^2_\psi$) is positive. In
order to make this scalar  healthy, the sound speed square must be
negative, but it is inevitably unstable. Thus, one way to avoid this
is to choose the case that the sound speed square is close to zero,
which implies the limit of $\lambda \to 1$. However, in this limit,
the cubic interactions are important at very low energies which
indicates the strong coupled problem~\cite{KA}. This invalidates any
linearized analysis and any predictability of quantum gravity is
lost due to unsuppressed loop corrections.  The authors~\cite{BPS2}
tried to extend the theory to make a healthy HL gravity, but there
has been some debate as to whether this theory is really
healthy~\cite{PS,BPS3,KP}.

 On the other hand, in  the Hamiltonian approach to the nonprojectable HL gravity, the
authors~\cite{LP} did not consider the Hamiltonian
 constraint as a second class constraint, which leads to a  strange
 result that there are no degrees of freedom left when  imposing the
 constraints of the theory. Moreover, the authors~\cite{HKG} have claimed that
 there are no solution of the lapse function which  satisfies  the
 constraints. Unfortunately, it implies  a surprising conclusion that there is no
 evolution  at all  for any observable.   However, more recently, it
 was shown that the IR version of HL gravity ($\lambda
 R$-model)  is completely equivalent to the general relativity for any
 $\lambda$  when employing a consistent Hamiltonian formalism based on Dirac algorithm~\cite{BR}.

The projectability condition from condensed matter physics may not
be appropriate for describing the (quantum) gravity. Instead, if one
does not impose the projectability condition, the HL gravity leads
to general relativity without the  strong coupling problem in the IR
limit. We note that  this work was carried out without the
projectability condition and thus, was nothing to do with the strong
coupling problem for $1/3 \le \lambda \le 1$.

\section*{Acknowledgement}
Y. Myung thanks Li-Ming Cao, Sinji Mukohyama, and Mu-In Park for
helpful discussions. H. Lee was supported by the National Research
Foundation of Korea (NRF) grant funded by the Korea government
(MEST) (No. 2009-0062869).
 Y. Kim was supported by the Korea Research Foundation Grant funded
by Korea Government (MOEHRD): KRF-2007-359-C00007.   Y. Myung  was
supported  by Basic Science Research Program through the National
Research  Foundation (NRF) of Korea funded by the Ministry of
Education, Science and Technology (2009-0086861).

\end{document}